\documentclass[conference]{IEEEtran}
\IEEEoverridecommandlockouts
\usepackage{cite}
\usepackage{amsmath,amssymb,amsfonts}
\usepackage{algorithmic}
\usepackage{hyperref}
\usepackage{graphicx}
\usepackage{textcomp}
\usepackage{xcolor}
\def\BibTeX{{\rm B\kern-.05em{\sc i\kern-.025em b}\kern-.08em
    T\kern-.1667em\lower.7ex\hbox{E}\kern-.125emX}}

\begin{document}

\title{Multi-Agent Discovery and Resource-Aware Autonomous Exploration of Scientific Datasets \\
}

\author{\IEEEauthorblockN{Aashish Panta}
\textit{University of Utah}\\
Salt Lake City, Utah, USA \\
aashish.panta@utah.edu
\and
\IEEEauthorblockN{Hugo Lee}
\textit{Jet Propulsion Laboratory, Caltech}\\
Pasadena, CA, USA \\
huikyo.lee@jpl.nasa.gov
\and
\IEEEauthorblockN{Giorgio Scorzelli}
\textit{University of Utah}\\
Salt Lake City, Utah, USA \\
scrgiorgio@gmail.com
\and
\IEEEauthorblockN{Kyongsik Yun}
\textit{Jet Propulsion Laboratory, Caltech}\\
Pasadena, CA, USA \\
kyongsik.yun@jpl.nasa.gov
\and
\IEEEauthorblockN{Valerio Pascucci}
\textit{University of Utah}\\
Salt Lake City, Utah, USA \\
valerio.pascucci@utah.edu}

\maketitle

\begin{abstract}
Modern scientific facilities and instruments generate datasets at scales that are difficult for individual researchers to discover, access, and explore. Although many datasets are publicly available, using them often requires familiarity with repository organization, data formats, multiresolution structures, and visualization parameters. We present WebVisus, a constrained and resource-aware multi-agent system for discovering and autonomously exploring remote, multiresolution scientific datasets. Given a natural-language research question, WebVisus identifies the user's intent and launches an autonomous exploration agent that examines slices, volumes, and timesteps while adapting data resolution and retrieval quality to available client memory and computational resources. This design supports progressive exploration without complete dataset downloads or manual configuration of low-level visualization parameters using natural languages. We report the system architecture, constrained agent protocol, resource-aware access mechanism, and case studies evaluating autonomous visual exploration and resource-aware agentic access across scientific datasets.

\end{abstract}


\section{Introduction}

Modern scientific instruments, high performance computing facilities, and imaging systems produce datasets that exceed the storage, transfer, and processing capacity of individual workstations. Climate simulations, biomedical images, materials scans, and other scientific collections contain petabytes of multidimensional data \cite{xue2021biomaterials,CHEN2026102905,11402408}. Progressive multiresolution methods \cite{Kumar2019,Usher2021,11044452} improve access by retrieving selected regions and levels of detail on demand \cite{10767643}. Efficient access alone, however, does not make these datasets easy to use.
For any analysis tasks, researchers may need to identify suitable data, interpret repository terminology, resolve dataset locations, select variables and timesteps, and configure visualization and transfer parameters. These tasks require knowledge of both the scientific domain and the underlying infrastructure. Those who understand the scientific objective but lack experience with a specific archive or visualization system may struggle to use available resources effectively. Reducing this barrier is important for democratizing AI-assisted science and enabling a broader scientific community to engage with large scale data \cite{hsu2021empowering,taufer2024enhancing}.

Existing systems address individual stages of this process. Scientific catalogs support metadata search and collection browsing \cite{zaidi2017data,subramaniam2021comprehensive}, but often assume familiarity with repository conventions. Progressive visualization frameworks provide interactive access after a dataset has been identified \cite{11297838}. Natural language interfaces translate user requests into analytical or visual specifications \cite{Tian_2025}, while agent based systems can manipulate visualization applications, optimize rendering parameters, and coordinate scientific workflows \cite{eliza2026animating}. Most of these systems begin after the dataset has been selected and made available to the analysis environment. The transition from an open ended scientific goal to grounded dataset discovery, feasible remote access, and iterative visual investigation under explicit resource constraints still remains unaddressed.

We address this gap through a multi-agent system, called WebVisus, for resource-aware autonomous exploration of remote scientific data. Given a natural language goal, the system interprets user intent, searches a catalog of approximately 1,400 datasets, identifies suitable collections using metadata grounded tools, and opens the selected data through a constrained action interface. The agent then explores slices, volumes, isovalues, viewing configurations, and timesteps while adapting data resolution and transfer size to memory and latency constraints. 
WebVisus supports autonomous exploration when datasets are too large to download or place directly within a model context. The agent reasons about both the scientific value of a view and the feasibility of retrieving and rendering it within the active resource budget. Dataset discovery, access planning, visualization control, and iterative evaluation form a single closed loop workflow. WebVisus automates the initial stages of data selection and visual investigation, helping researchers reach informative views with less repository-specific and interface-specific expertise while preserving human control over scientific interpretation.

This paper makes three contributions:

\begin{itemize}
    \item An end-to-end multi-agent workflow that connects natural-language scientific goals to metadata-grounded discovery and selection over approximately 1,400 datasets, with the ability to add new datasets on-the-fly.
    \item A bounded autonomous exploration agent that iteratively observes the current analysis state, selects visualization operations, evaluates the resulting view, and records its actions through an inspectable execution protocol.
    \item A agent-enabled resource-aware mechanism that autonomously estimates the cost of data requests and adapts resolution, access mode, spatial extent, and transfer size to runtime constraints.
\end{itemize}

 \vspace{-1mm}

\begin{figure*}
    \centering
    \includegraphics[width=0.89\linewidth]{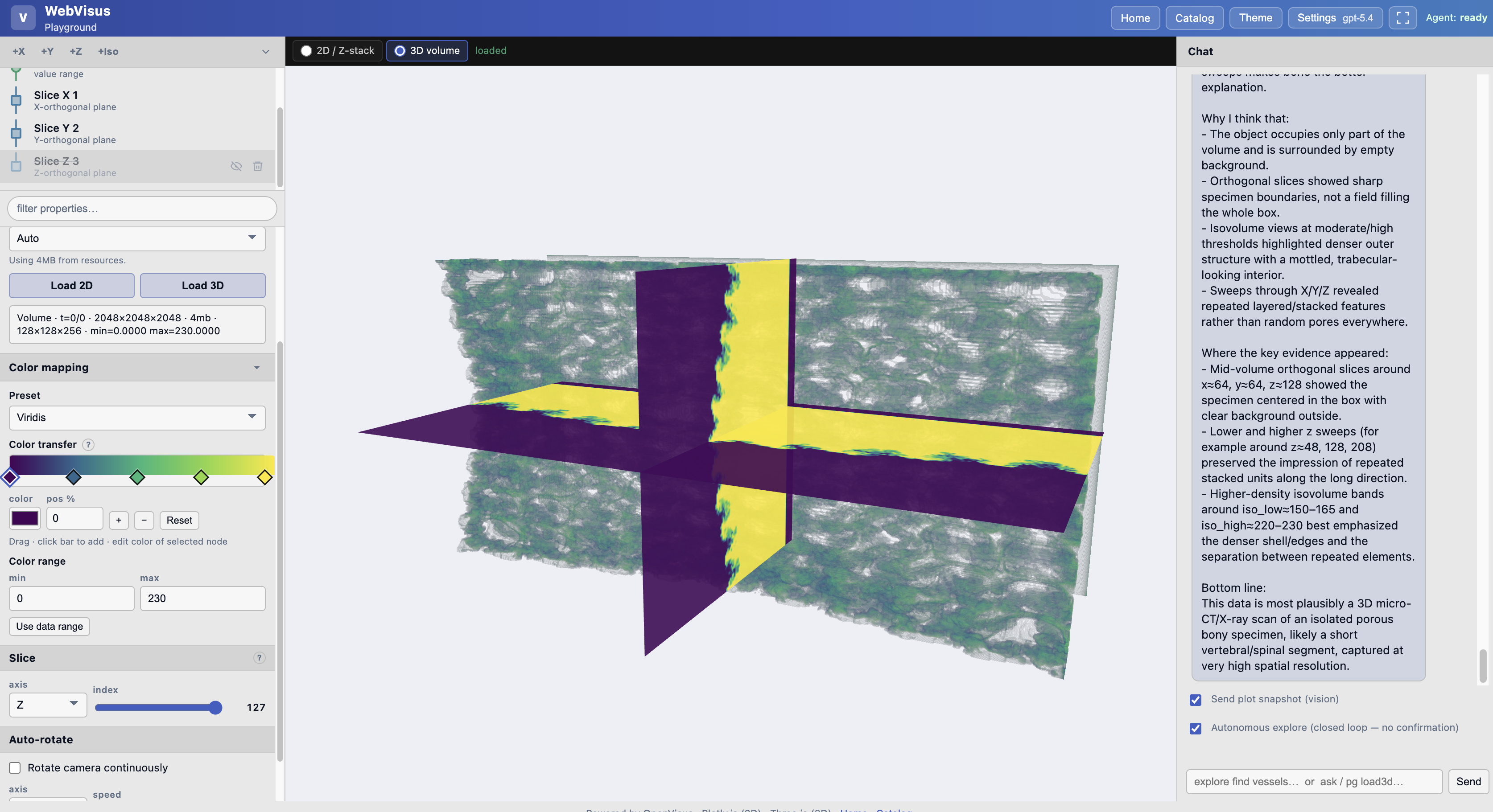}
    \caption{WebVisus interface for interactive and autonomous exploration of large volumetric data. The left panel provides
manual control over slices, transfer functions, isovalues, rendering quality, and camera motion; the center displays the
resulting 3D volume and orthogonal slice views; and the right panel shows the agent’s observations and findings. During
autonomous exploration, the agent iteratively modifies slices along all three axes, adjusts isovalue and rendering parameters,
rotates the view, and evaluates the resulting visual evidence. These operations are supported by resource-aware progressive
retrieval, which fetches only the data resolution and region required for each analysis step.}
    \label{fig:UI}
\end{figure*}

\section{Related Work}

Large-scale scientific visualization and intelligent support for high-performance computing have been studied for many years. Recent advances in large language models have expanded these areas through natural-language interaction, tool use, and autonomous scientific workflows. We review prior work in two areas that are most relevant to WebVisus, namely LLM-based tools for scientific data visualization and agentic AI for HPC and autonomous science.

\subsection{LLM-Based Tools for Scientific Data Visualization}

Natural language has long been studied as an interaction modality for visualization systems because it allows users to express analytical intent without directly manipulating complex interfaces~\cite{9699035,fu2020quda,10026499}. Early systems such as NL4DV demonstrated that natural-language queries could be translated into analytical specifications and basic visualizations~\cite{narechania2020nl4dv}. Recent work has incorporated large language models into visualization design and visual analysis as well. These systems use LLMs to generate design feedback, improve visualization quality, support analytical reasoning, and assist users in operating complex tools~\cite{shin2025visualizationary,misvisfix,zhao2024leva,gao2024fine,Tian_2025}. LLM-based methods have also been applied to various domain-specific visual-analysis workflows such as climate and clinical stroke-data analysis~\cite{rashik2025claimate,kim2024phenoflowhumanllmdrivenvisual}.

Scientific visualization presents additional challenges because datasets are often volumetric, time varying, and too large to be included directly in a model context\cite{9830790,sun2023scalable}. AVA~\cite{ava_liu_2024} showed that language models can act as autonomous visualization agents for tasks such as transfer-function optimization and dimensionality-reduction analysis. NLI4VolVis~\cite{ai2025nli4volvis} uses LLM agents to explore and edit scientific scenes represented through 3D Gaussian splatting. ParaViewMCP~\cite{liu2025paraview} connects a language model to ParaView through the Model Context Protocol, enabling natural-language control of scientific visualization workflows. InferA~\cite{10.1145/3731599.3767342} coordinates multiple agents to analyze large and complex scientific simulation data. These studies demonstrate the potential of language models for visualization assistance, interface control, and automated analysis. Many existing approaches, however, assume that the dataset has already been selected, loaded, or reduced to a manageable representation~\cite{narechania2020nl4dv,gao2024fine,liu2025paraview}. This assumption is difficult to maintain for large scientific volumes and time-varying simulation datasets.

WebVisus addresses this gap through a closed-loop workflow over remote multiresolution data. The agent observes the current visualization, selects subsequent exploration actions, and adjusts resolution and transfer size according to computational constraints. This supports language-driven visual exploration of remote scientific datasets even when the complete dataset cannot be downloaded on the client side.

\subsection{Agentic AI for HPC and Autonomous Science}
Recent surveys describe agentic AI for science as systems that combine reasoning, planning, and autonomous decision-making to support literature review, hypothesis generation, experimentation, and result analysis~\cite{gridach2025agentic}. Zhou et al.~\cite{zhou2025autonomous} describe scientific agents as autonomous systems that coordinate human input, natural language, and physical models across the data discovery process. Agent Laboratory demonstrates how LLM agents can support an end-to-end research workflow by conducting literature review, experimentation, and report writing from a human-provided idea, while allowing researchers to provide feedback and guidance throughout the process~\cite{schmidgall2025agent}.
Ward et al. present an agent-based framework that adapts scientific workflows during execution on supercomputers, including applications in chemistry, materials science, and biophysics\cite{ward2025employing}. Ma et al. connect LangChain and LangGraph agents to Parsl, allowing LLM-generated tool calls to execute concurrently on the Polaris supercomputer \cite{ma2025connecting}.
Kamatar et al. introduce Academy\cite{kamatar2026empowering}, which deploys coordinated agents across HPC systems, experimental facilities, and data repositories. Lu et al. present a system that generates research ideas, writes code, runs experiments, analyzes results, and produces scientific papers\cite{lu2024ai}.

These studies establish the potential of agents for autonomous exploration and scientific visualization, but they largely begin after dataset selection and access have been resolved. WebVisus extends the agentic workflow to include metadata-grounded discovery, resource-feasible remote access, and bounded visual exploration. Next, we describe the system architecture that connects these stages.

\begin{figure*}[t]
    \centering
       
    \includegraphics[width=0.84\linewidth]{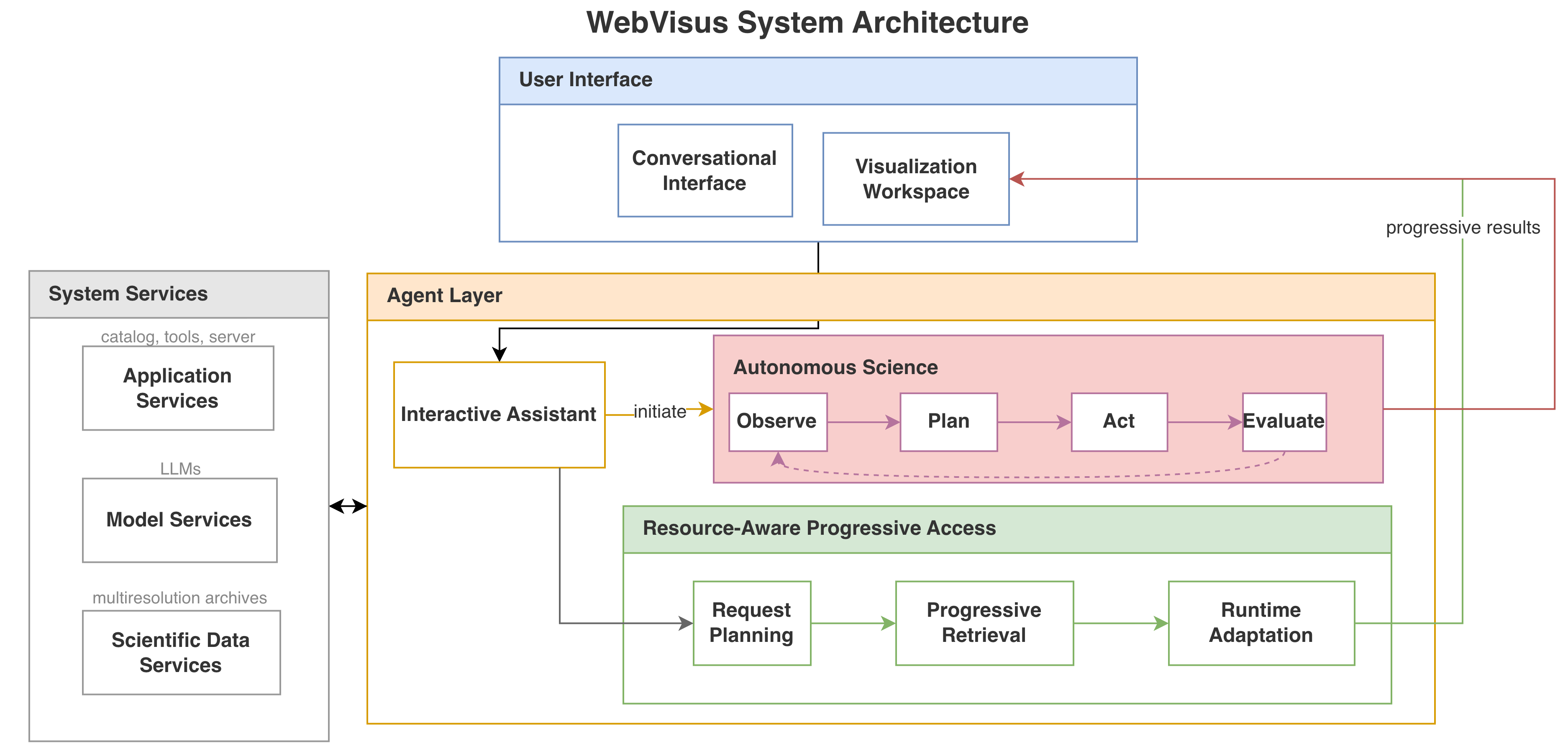}
    \caption{WebVisus system architecture. Users explore scientific volumes through interactive and autonomous operating modes that share a three-agent runtime consisting of an Orchestrator, Catalog Agent, and Executor. The interactive assistant handles turn-by-turn requests and can initiate an autonomous science loop that repeatedly observes the current analysis state, plans and executes visualization operations, and evaluates the resulting evidence. Analysis requests are mediated by a resource-aware progressive-access layer that plans data requests, incrementally retrieves multiresolution data, and adapts execution to available computational resources before returning progressive results to the visualization workspace. Application, model, and scientific data services provide system functionality, language-model inference, and access to multiresolution data archives, respectively.}
    \label{fig:overview}
    
\end{figure*}
 \vspace{-1.5mm}

\section{System Architecture}
\label{sec:architecture}

WebVisus is a multi-agent system for autonomous exploration of remote,
multiresolution scientific data. 
Its architecture consists of three control layers: a browser client, an
application server, and a cloud-hosted language-model service. The browser
provides the Catalog and Playground interfaces. The application server
maintains the catalog index,  validates agent actions, and
enforces resource limits. Language-model inference is provided by an Azure
OpenAI deployment in Microsoft Foundry\cite{microsoft_foundry}. The current
deployment provides access to multiple model backends, including
DeepSeek-V4-Flash, Grok-4.3, Kimi-K2.6, GPT-5.4-Pro, GPT-5.4, and o3-mini;
multimodal vision input is supported by a subset of these models. Scientific data remain on the remote Atlantis 
service at The Center For High Performance Computing(CHPC)\cite{utah_chpc}, and we only retrieve the resolution, region, and timestep
required by the current operation. 
This separation allows autonomous operation without granting a language model
unrestricted access to the browser, operating system, or data service.
Reasoning is delegated to the model backend, whereas executable operations
remain controlled by deterministic application components. The following
subsections describe the catalog-grounded multi-agent runtime, action
protocol, progressive-access model, visualization pipeline, and autonomous
exploration loop.

\subsection{Catalog-Grounded Multi-Agent Runtime}
\label{sec:catalog-indexing}

WebVisus maintains a catalog snapshot containing dataset names, group paths,
nested entries, and temporal information derived from the Atlantis server
described earlier. The current snapshot indexes approximately 1,400 datasets
across more than 40 groups, including climate simulations, biomedical volumes,
materials science products, microscopy data, and other scientific collections
reaching terabytes in size.  We use token
normalization and  semantic search to allow conversational requests to be matched with the data even if users don't
know the exact dataset name or repository terminology.

Natural-language requests are processed through two operating modes, interactive assistance and autonomous exploration, that share the same three specialized backend agents: an \emph{Orchestrator}, a \emph{Catalog Agent}, and an \emph{Executor}. An \emph{Orchestrator} interprets
the user's objective and decomposes it into a structured plan. The plan
specifies whether catalog evidence is needed, which visual target is implied
(e.g., overview, slice, volume, or temporal inspection), and whether the
request should prioritize exploration breadth or resource feasibility. The
Orchestrator does not directly query the catalog or issue visualization
commands, but delegates those tasks to downstream agents and revises
the plan based on their outputs.

When catalog grounding is required, a \emph{Catalog Agent} independently
invokes a fixed set of tools for retrieving catalog statistics, listing
groups, searching metadata, recommending relevant datasets, and resolving a
dataset name to its location. It returns the best match, ranked alternatives,
and supporting metadata for access planning. This agent operates as the
specialist for dataset discovery and repository grounding.

An \emph{Executor} receives the Orchestrator's plan, the Catalog Agent's
results, and the current interface state, then translates them into typed
actions for the runtime. These actions may select a dataset, choose an access
mode, set a timestep, adjust resolution, or modify the visualization. The
Executor also emits intermediate state updates that can trigger another round
of Orchestrator reasoning when the current representation is insufficient.

\subsection{Constrained Action Protocol}
\label{sec:action-protocol}

WebVisus restricts agent execution to a registered vocabulary of typed
commands. The language model cannot directly manipulate the document object
model, invoke a shell, access the operating system, or construct arbitrary
requests to Atlantis. It proposes operations that must be parsed and matched
to handlers implemented by the application. The Catalog interface accepts commands for search, preview, and dataset
selection. The Playground accepts commands for loading data, selecting
quality and access mode, changing timesteps, applying colormaps, placing
orthogonal slices, adding isovolumes, modifying opacity, controlling the
view and modifying other visual parameters. Commands outside the registry, with invalid parameters, or violating application-defined resource and safety constraints are rejected deterministically.

Each Playground request may include a compact representation of the current
state, including the selected dataset, access mode, timestep, quality level,
and active visualization nodes. When multimodal reasoning is enabled, a
snapshot of the rendered view is also attached. The model therefore reasons
from the current application state instead of reconstructing it only from
conversation history.
Successful and failed operations are appended to the interaction transcript.
This record makes agent behavior observable and allows users to inspect the
sequence of visualization and resource decisions. The model determines what
operation to propose, but the executable surface remains fixed, bounded, and
auditable.

\subsection{Resource-Aware Progressive Access}
\label{sec:resource-access}

Autonomous science over large data requires an agent to reason about both
scientific relevance and computational feasibility. WebVisus accesses
Atlantis datasets through progressive reads, allowing each
request to specify a resolution, spatial selection, access mode, and, where
available, a timestep~\cite{Kumar2019,summa2011interactive}. Instead of treating a complete
dataset as a single unit of analysis, the system selects a representation that
fits the available memory and transfer budget.

\subsubsection{Memory-Aware Resolution Selection}
\label{sec:multires}

Because the data are already stored as arrays in a multiresolution layout, WebVisus can
estimate the decoded memory footprint of a candidate representation before
loading it. The system can then select the finest available resolution that
satisfies a prescribed memory budget without requiring the user or agent
to guess which representation is feasible.

We describe a resolution using an axis-refinement string
$s \in \{0,1,2\}^{*}$. Each symbol indicates the spatial axis refined at that
step: $0$, $1$, and $2$ correspond to the $X$, $Y$, and $Z$ dimensions,
respectively. Let $n_0(s)$, $n_1(s)$, and $n_2(s)$ denote the number of
refinements along the three axes. The resulting array shape is

\[
R(s)=
\left(
2^{n_0(s)},
2^{n_1(s)},
2^{n_2(s)}
\right).
\]

For an array containing $b$ bytes per sample, the estimated decoded memory
footprint is

\[
M(s)
=
b\prod_{i=0}^{2}2^{n_i(s)}
=
b\cdot2^{n_0(s)+n_1(s)+n_2(s)}.
\]

Consider the refinement string

\[
s=010101201201201012012012012012012.
\]

It contains

\[
n_0(s)=12,\qquad
n_1(s)=12,\qquad
n_2(s)=9,
\]

which produces

\[
R(s)=
\left(
2^{12},
2^{12},
2^{9}
\right)
=
(4096,4096,512).
\]

For \texttt{float32} data, where $b=4$ bytes, the decoded array requires

\[
\begin{aligned}
M(s)
&=4\cdot2^{12}\cdot2^{12}\cdot2^{9}\\
&=34{,}359{,}738{,}368\text{ bytes}=32\text{ GiB}\\
\end{aligned}
\]

This representation exceeds a memory budget of 4~GiB. WebVisus can instead
select a shorter refinement string,
\[
s'=010101201201201012012012012012,
\]
for which the resulting shape is
\[
R(s')=
\left(
2^{11},
2^{11},
2^{8}
\right)
=
(2048,2048,256),
\]
with decoded footprint
\[
\begin{aligned}
M(s')
&=4\cdot2^{11}\cdot2^{11}\cdot2^{8}\\
&=4{,}294{,}967{,}296\text{ bytes}=4\text{ GiB}\\
\end{aligned}
\]

Given a memory budget $B$, WebVisus selects the finest candidate refinement
string satisfying

\[
M(s)\leq B.
\]

\subsubsection{Agentic Resource Control}

Resolution selection is exposed as an explicit agent decision. Using the Orchestrator's interpretation of the
scientific objective, the Executor chooses an initial quality, 
spatial extent, and timestep. The estimated footprint of this
data is then compared with the active budget.
If a request exceeds the budget, WebVisus can lower the quality, restrict the
region, select a slice instead of a volume, or limit temporal access to fixed timesteps until the estimated array footprint is feasible(equal or lower).  After the read completes, the agent receives the
final array shape, loading duration, and current visualization state. It
may increase detail when the current representation is insufficient or reduce
the request when memory use or latency becomes excessive.

The server also checks the data size before sending it to the browser. The first resource check limits how much data is loaded into server memory, while this second check limits how much data is transferred to and displayed in the browser. If the result is too large, WebVisus can reduce the resolution, select a smaller region, or return a slice instead of the full volume. This helps reduce transfer time, browser memory use, and rendering cost. These decisions are shown to the user and can be manually changed.
 Resource decisions and their execution results remain
visible in the transcript and may be overridden by the user.

\subsection{Visualization Pipeline}

After data retrieval, the Playground constructs an interactive visualization
from the returned scalar field and metadata. Two-dimensional arrays are
displayed using Plotly.js ~\cite{plotly}, while
three-dimensional arrays are uploaded to a client-side WebGL pipeline
implemented with Three.js~\cite{threejs}. A ParaView-style pipeline
organizes the volume source, orthogonal slices, isovolume bands, transfer
functions, opacity, and camera controls~\cite{ahrens2005paraview}. These
operations sample the same cached texture, allowing slice positions,
colormaps, opacity, isovalue ranges, and camera settings to change without
additional data retrieval. New server requests are required only when the dataset, resolution, spatial
selection, access mode, or timestep changes. This separation allows
progressive access to determine how much data enters the browser while the
client-side pipeline determines how that representation is examined.

\begin{figure*}[t]
    \centering
    \includegraphics[width=0.85\linewidth]
    {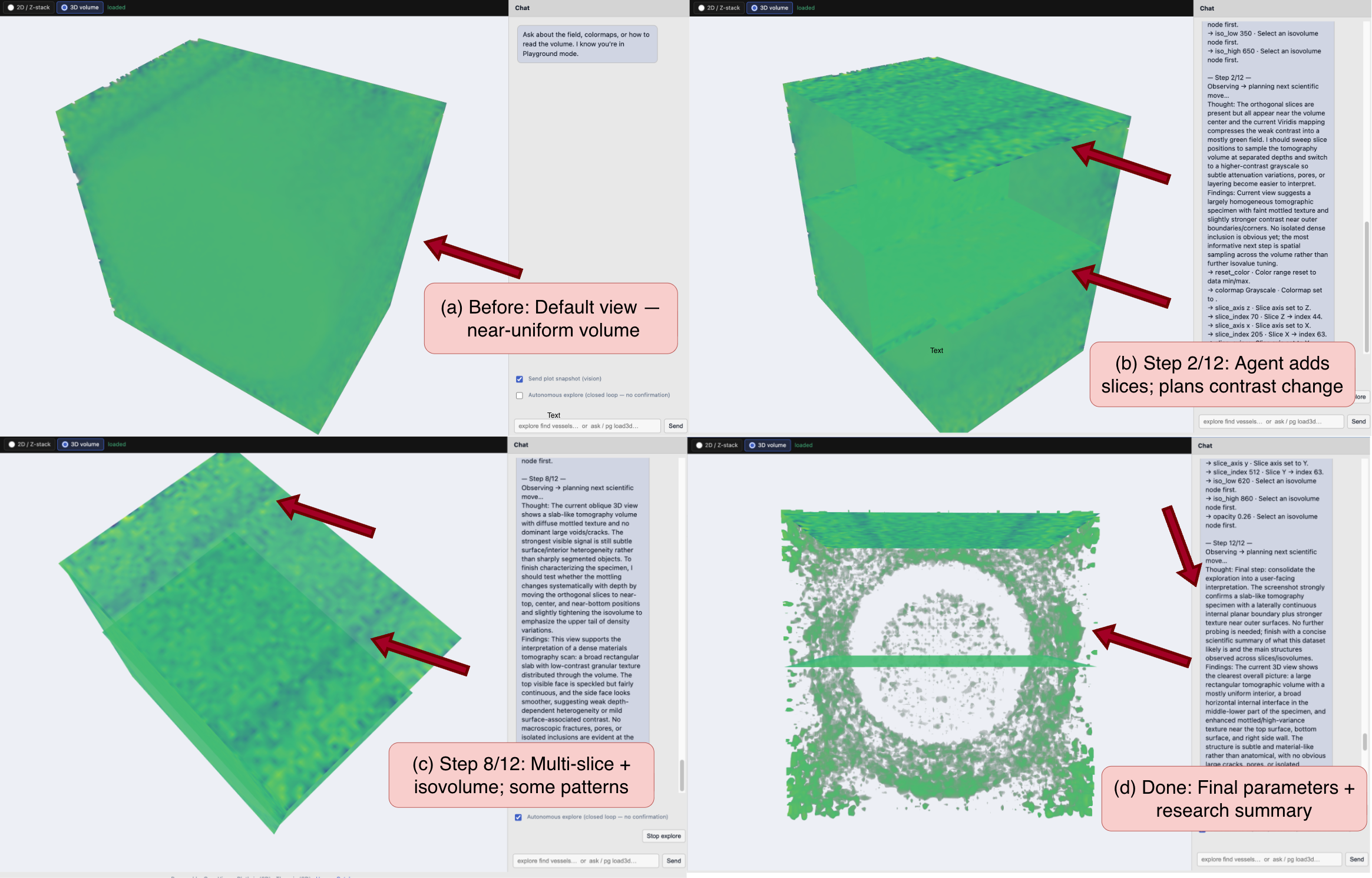}
    \caption{Autonomous exploration of a materials-science tomography volume.
    (a) The default rendering shows little internal contrast.
    (b) At step 2 of 12, the agent adds orthogonal slices and begins adjusting
    the visual representation.
    (c) At step 8, combined slices and an isovolume reveal spatial variation
    that is not apparent in the initial view.
    (d) At step 12, adjusted opacity, isovalue, and slice positions expose a
    candidate internal boundary and support a final exploration summary.
    No intermediate visualization parameters were selected by the user.}   
    \vspace{-4mm}

    \label{fig:auto-arr}

\end{figure*}

\subsection{Autonomous Exploration Agent}
\label{sec:autonomous}

The Autonomous Exploration Agent combines catalog grounding, progressive
access, and constrained visualization control in an
Observe-Plan-Act-Evaluate loop. A researcher provides a natural-language
objective, such as locating internal structure in a tomography volume or
examining selected states of a climate simulation. WebVisus resolves the
relevant dataset, retrieves an initial representation within the resource
budget, and allows the agent to determine the subsequent operations on its own.

At each iteration, the system observes the current Playground state. The
observation may include the loaded array shape, value statistics, selected
parameters,  recent operations, and a snapshot of the
rendered view. The model receives this information together with the
scientific objective and relevant catalog metadata. It then  evaluates the current state, and proposes a small set of actions
towards the goal. Changes to resolution, spatial extent, access mode, or
time are evaluated by the resource policy. Changes to slices, isovalues,
opacity, colormap, or camera state operate on the cached representation. After
an action that triggers data retrieval, the controller waits until loading
and rendering complete before beginning the next iteration.

The agent may revise its strategy when a view is uninformative or an
operation is too expensive. It can inspect another axis, switch between slice
and volume views, move through time, adjust transfer functions, or reduce the
requested representation when latency or memory cost becomes excessive. A
client-side backoff mechanism further lowers retrieval settings when an
operation times out or exceeds the expected duration.

Execution stops when the objective is judged complete, a fixed step limit is
reached, or the user issues a stop command. WebVisus then produces a summary
grounded in the sequence of observations, interface states, and rendered
views encountered during the run. The transcript records the observations,
plans, actions, execution results, and resource decisions that contributed to
the final response.
This design extends autonomous science beyond agents that manipulate a single
preloaded dataset. WebVisus can operate across
datasets autonomously, determine which collection is relevant, select a computationally feasible representation, and adapt its investigation from intermediate
observations. 



\section{Evaluation}
\label{sec:evaluation}

We evaluate WebVisus through two studies. The first examines
whether an agent can autonomously improve an initially uninformative
visualization of a remote tomography volume while remaining within a fixed
resource budget and constrained action space. The second measures progressive
access latency across six datasets to determine which data scales support
closed-loop autonomous exploration.

 \vspace{-1.5mm}
\subsection{Case Study 1: Autonomous Exploration of Tomography Data}
\label{sec:case-arr}

\subsubsection{Objective and Setup}

This study evaluates whether WebVisus can improve an initially uninformative rendering of a large remote volume without intermediate user intervention while enforcing the resource limits.
The experiment used a single timestep tomography volume with a logical shape of approximately $707 \times 2048 \times 2048$. A fully decoded \texttt{float32} volume would require approximately 11~GiB. Also, the generic dataset name provided limited information about the specimen or its expected structures.

The user opened the dataset and requested identification of potentially interesting internal structures. WebVisus retrieved a progressive representation with an approximate shape of $45 \times 64 \times 64$. The user made no subsequent interface changes. The run was limited to 12 Observe, Plan, Act, and Evaluate iterations. At each step, the agent received the current visualization, array statistics, application state, and recent action history, while also increasing the resolution as needed.

\subsubsection{Autonomous Exploration Behavior}

Figure~\ref{fig:auto-arr}a shows the initial rendering, which appears as a nearly uniform slab with limited contrast. The agent treated this view as inconclusive and evaluated alternative slices, isovalues, opacity settings, viewing directions, and representations.

During the early iterations, the agent added orthogonal slices along the $X$, $Y$, and $Z$ axes. At step 2, shown in Figure~\ref{fig:auto-arr}b, it detected weak contrast and adjusted the slice positions and appearance parameters.
By step 8, the agent had combined multiple slices with an isovolume, as shown in Figure~\ref{fig:auto-arr}c. This configuration revealed spatial variation that was difficult to distinguish initially. Further adjustments to opacity, isovalue ranges, and slice locations produced the final view in Figure~\ref{fig:auto-arr}d, which exposed a cylindrical internal feature. WebVisus then summarized the visited states, parameter changes, and candidate observations.
Using GPT-5.4 as a default model for all agents, the complete workflow required 172 seconds and approximately 94,000 tokens, including dataset inspection, parameter manipulation, and summary generation.

The observations remain exploratory and require domain validation. The agent located informative views and candidate structures for subsequent analysis by a specialist. Direct quantitative analysis through the interface remains future work.

\subsubsection{Resource Constraints and Oversight}

The full resolution volume remained on Atlantis throughout the run. WebVisus retrieved only the progressive representation required for each operation and completed every data load before capturing the next observation.
 When an operation exceeded the expected memory or latency budget, the controller could reduce the quality, transfer size, or spatial extent. The agent selected the inspection target, while WebVisus enforced retrieval and rendering limits.
The user could interrupt execution at any time. The interface recorded all commands, results, and observations, and rejected unregistered actions. The agent had no direct access to the browser document model, operating system, or shell.
The trace also exposed an action dependency limitation. Several early isovalue adjustments failed because the required isovolume node had not been created. The agent continued slice based inspection and later added the node. This recovery preserved task completion and motivates explicit command precondition validation.

\subsubsection{Result}

The study demonstrates first pass autonomous investigation of a large remote dataset. The resource aware agent revised its strategy, identified a candidate structure, and generated a traceable summary without user control of intermediate visualization parameters. The final configuration exposed spatial variation that was unclear in the initial rendering and produced an auditable exploration trace of its results.

 \vspace{-1mm}
\subsection{Case Study 2: Resource-aware Exploration at  Scale}
\label{sec:case-progressive}

This study evaluates the access conditions required for autonomous exploration of large remote archives. We measure how retrieval latency changes as WebVisus increases the amount of progressive data available to the agent.

\subsubsection{Experimental Method}

We measured remote access across six datasets representing tomography, atmospheric simulation, micro CT, microscopy, agriculture, and materials science. Each time measurement here includes cloud retrieval and local loading time.
For small and moderate presets, WebVisus selected a progressive resolution corresponding to the requested data size. Larger presets used additional regions, slabs, or timesteps to approximate the target volume. Smaller representations reduce retrieval, memory, and rendering costs but may omit fine scale structures. Larger representations provide more detail while increasing processing, transfer, memory, and rendering costs. Autonomous exploration must therefore balance visual information with responsiveness.

\begin{figure}[t]
\centering
\includegraphics[width=\columnwidth]
{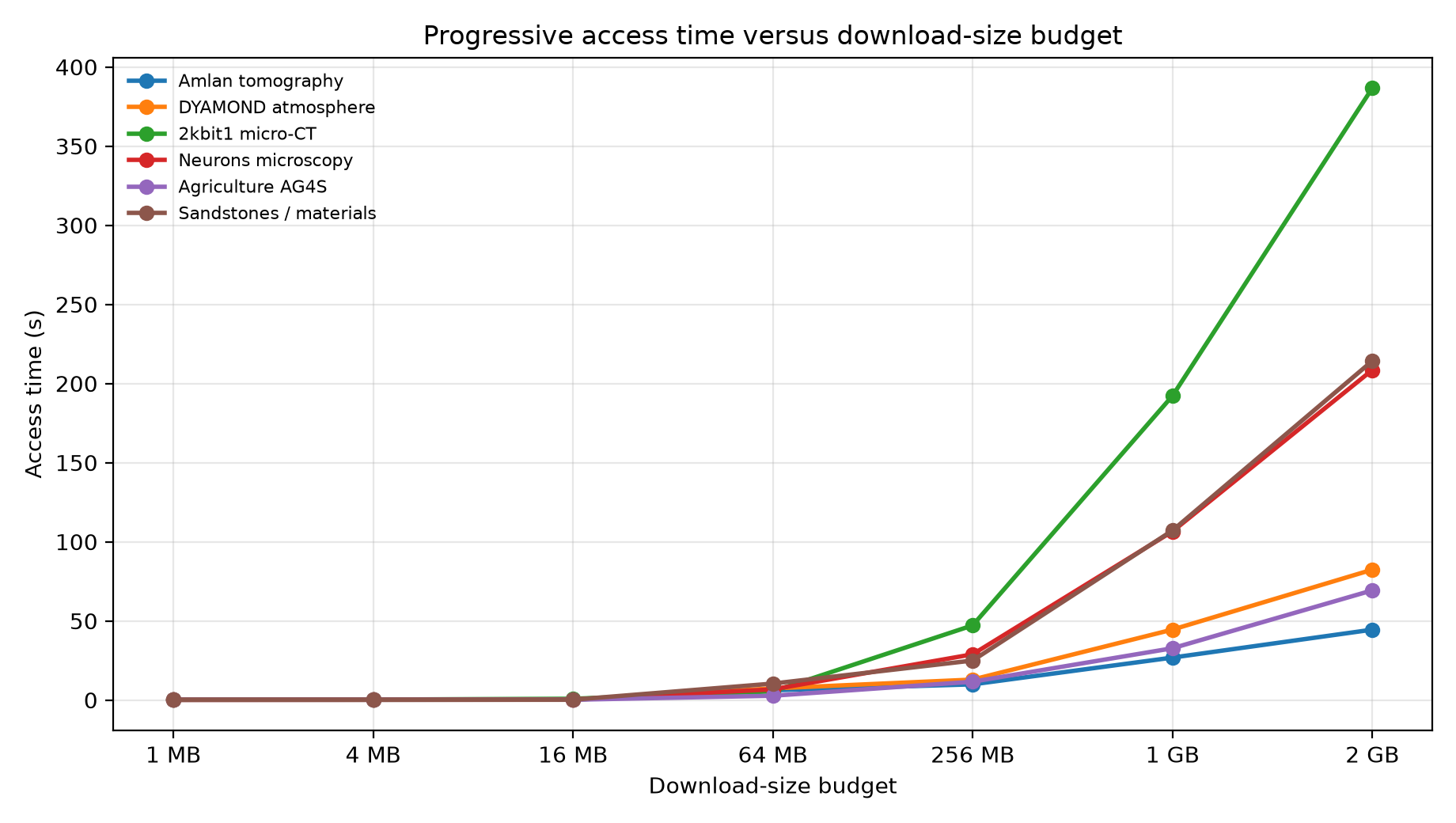}
\caption{Progressive access time across six scientific datasets. Megabyte scale representations support repeated observations, while larger transfers introduce substantial and dataset dependent delays. WebVisus uses measured latency and estimated memory cost to guide subsequent access decisions.}
\label{fig:size_v_time}
\end{figure}
\subsubsection{Measured Access Behavior}

Figure~\ref{fig:size_v_time} reports cloud retrieval and local loading time as the requested data size increases. At 1MB, the tested datasets load in approximately 0.1-0.2 s, allowing repeated retrieval without dominating model inference, action execution, and rendering time.

Latency remains low across the megabyte range but increases sharply for larger requests. At 256MB, access times range from approximately 10 to 47s. Requests of 1 GB and 2 GB data require tens to hundreds of seconds. At 2GB, retrieval reaches approximately 387s for the \texttt{2kbit1} micro CT dataset and exceeds 200s for the neurons and sandstone datasets.
The measurements identify two patterns here. Smaller scale representations support repeated observation and action, while larger transfers can stall autonomous execution. The largest feasible request may therefore be unsuitable for iterative analysis all the time.

\subsubsection{Adaptive Agent Policy}

WebVisus includes transfer size and resolution in the agent action space. Before retrieval, the system considers the logical dimensions, field type, timestep range, memory budget, and transfer preset. For fast 3D interactivity, initial requests commonly range from 512KB to 2MB, depending on the dataset and objective.
After retrieval, the agent receives the returned shape, estimated decoded memory, measured loading time, visualization state, and rendered image. It may increase the resolution, restrict the spatial region, inspect another slice or slab, change between two and three dimensional access, or select another timestep.

The agent retains the current representation when it provides sufficient information under the available resource budget. Ambiguous views can trigger higher resolution access or sampling at another spatial or temporal location. WebVisus reduces the requested size or quality when retrieval latency or memory use exceeds the active policy.
Every resource decision uses the registered action protocol and appears in the exploration trace. The researcher can inspect these decisions, override the selected parameters, or stop execution.

\begin{table*}[t]
\centering
\caption{Outcomes of five autonomous exploration workflows.}
\label{tab:autonomous-workflows}
\scriptsize
\setlength{\tabcolsep}{3pt}
\renewcommand{\arraystretch}{1.15}
\begin{tabular}{
    c
    p{3.0cm}
    p{3.0cm}
    p{6.2cm}
    r
    r
    r
    c
}
\hline
\textbf{ID} &
\textbf{Objective} &
\textbf{Dataset} &
\textbf{Agent Output} &
\textbf{Time} &
\textbf{Tokens} &
\textbf{Outcome} \\
\hline

T1 &
Find interesting patterns in the current data &
\texttt{amlan-tomography-1} &
Isovolume and orthogonal slices revealed an elliptical ring, prism shaped geometry, and granular alignments &
172\,s &
90\,k &
Success \\

T2 &
Find a three dimensional tomography or micro CT dataset and inspect its internal structure within 2\,MB &
\texttt{4D\_micro\_CT\_sandstones} &
A central $Z$ slice and isovolume revealed high attenuation grains within a lower density matrix and visible pore regions &
25\,s &
12\,k &
Success \\

T3 &
Identify the current dataset and produce an internal overview &
\texttt{Tomography} &
Multiple slices and an isovolume revealed concentric radial bands and granular texture &
99\,s &
52\,k &
Success \\

T4 &
Characterize pore and solid regions using slices and an isovolume &
\texttt{2kbit1} &
Multiple isovalue bands and orthogonal slices revealed pore regions within a connected solid matrix &
153\,s &
88\,k &
Success \\

T5 &
Find a time varying atmospheric volume and inspect its spatial and temporal structure within 2\,MB &
\texttt{NASA DYAMOND Data} &
Four temporal samples with slices and an isovolume revealed evolving convective and frontal structures &
89\,s &
51\,k &
Success \\
\hline
\end{tabular}
\vspace{-3mm}
\end{table*}

 \vspace{-1.5mm}
\subsection{Autonomous Exploration workflows}
\label{sec}

\subsubsection{Protocol and Success Criteria}

Here, we report five Autonomous Explore workflow using multiresolution datasets in the Table \ref{tab:autonomous-workflows}. Each workflows used vision-enabled plot snapshots, a maximum of 12 iterations, and no intermediate user actions throughout the analysis. The user provided a single natural language objective, after which the agent selected or used a dataset, retrieved a progressive representation, applied registered actions, and produced an exploration summary. 
A workflow was considered successful when WebVisus loaded a dataset, executed the list of predefined tasks, produced a nonempty summary, and terminated  without user interruption.

\subsubsection{Results}

All five workflows satisfied the success criteria without intermediate user actions. The workflows used an average of 7.8 of the 12 available iterations and required an average end to end time of approximately 108 seconds. The agent exercised orthogonal slicing, isovolume construction, appearance controls, spatial sampling, and temporal sampling across the evaluated datasets.

Across the five workflows, the agent proposed 203 Playground actions. The command layer executed 175 actions successfully, corresponding to $86.20\%$, and rejected 28 actions($13.8\%$). Rejections commonly occurred when the agent attempted to modify a parameter before creating the associated visualization node. Despite these failures, the agent recovered and completed every workflows. These results demonstrate bounded task completion while motivating explicit command precondition checks and stronger action recovery policies.
The generated summaries provide exploratory first contact assessments intended to guide subsequent expert analysis. They represent  structures and observations in the data without establishing validated scientific conclusions.

\subsubsection{Implications for Autonomous Science}

The measurements show that progressive access alone does not guarantee
interactive autonomous operation. The selected representation must reflect
both the computational environment and the information required at the
current step. A fixed high-res input  introduces long delays and may
exceed memory limits, while a permanently coarse policy may not reveal relevant
features.

We address this tradeoff by making data access part of agent
reasoning. The agent selects the resolution, spatial extent, access mode, and
timestep used to construct each observation, then considers both the visual
result and the cost of obtaining it before selecting the next action. This
extends agentic control beyond manipulation of an already loaded
visualization.
The tomography example in Section~\ref{sec:case-arr} illustrates the resulting
behavior. A progressive representation retrieved in approximately 0.16\,s was
sufficient to reject the default view, inspect multiple axes, adjust
isovolume parameters, and identify internal spatial variation without
downloading the approximately 11~GB logical volume locally. 

These results support the feasibility of resource-aware autonomous
exploration over a heterogeneous scientific archive. They do not yet measure
scientific correctness, information gain, or repeatability across agent runs.
Future evaluation will compare adaptive access against fixed-resolution
policies and report repeated-trial statistics, action-validity rates,
resource-policy compliance, end-to-end latency, and agreement with
domain-expert assessments.

 \vspace{-2mm}
\section{Discussion and Limitations}

WebVisus demonstrates a practical form of bounded autonomous science for large remote datasets. Through natural-language interaction, the agent can discover and open data, inspect slices and volumes, adjust isovalues and viewing parameters, and move across timesteps. Execution remains restricted to registered interface actions, allowing researchers to inspect, interrupt, or override each step.

Resource awareness is essential to this workflow. Browser-based analysis is constrained by network transfer, client and graphics memory, and server-side read costs. WebVisus estimates the decoded footprint $M(s)$ before retrieval and adapts the resolution, spatial extent, access mode, or timestep when a request exceeds the active budget. The agent therefore reasons jointly about what information may be useful and what representation can be processed safely. By combining this capability with catalog-grounded discovery, WebVisus lowers the expertise required to work with multiresolution archives and specialized visualization systems, making AI-assisted scientific exploration more accessible to domain scientists, students, and collaborators.

The current system remains limited to exploratory navigation and visualization. It does not perform validated scientific measurements, and the present evaluation does not establish repeatability, scientific correctness, or agreement with domain experts yet. Future work should add derived measurements, temporal comparison, feature detection, explicit stopping criteria, and provenance records. Future evaluation should also include repeated runs, fixed-resolution baselines,  exploration coverage, and end-to-end latency. These extensions are necessary to move from auditable first-pass visual exploration towards more reproducible autonomous scientific analysis.

 \vspace{-2mm}
\section{Conclusion}
We presented WebVisus, a browser-based system for autonomous exploration of remote, multiresolution scientific datasets. WebVisus combines an Observe-Plan-Act-Evaluate loop with resource-aware progressive access, estimating the memory footprint of each candidate representation before retrieval and adjusting resolution, spatial extent, and timestep to fit available memory and latency. This lets the agent explore terabyte-scale volumes without full downloads or manual tuning of low-level visualization parameters. Catalog-based discovery across roughly 1,400 datasets provides the starting point for this process, and all retrieval and rendering decisions remain bounded by explicit resource limits and recorded in an inspectable transcript. The system shows that autonomous visual exploration can be extended to large remote archives while keeping resource use adaptive and under human oversight.

\section*{Acknowledgment}
This work was funded in part by NSF awards 2609465, 2138811,  2127548, 2330582, the Advanced Research Projects Agency for Health (ARPA-H) grant no. D24AC00338-00.This work was performed in part under the auspices of the DoE by LLNL under contract DE-AC52-07NA27344 (LDRD project 25-SI-006).
HL and KS’s contribution to this work was performed at the Jet Propulsion Laboratory, California Institute of Technology, under a contract with NASA. We thank NASA for supporting the Regional Climate Model Evaluation System project. We thank Juan Carlos and Chris Woodland from Microsoft, and NASA SMCE for their support. GitHub Copilot was used in part for managing repository and some code support.

\bibliographystyle{acm}

\bibliography{template}

\end{document}